\documentclass[a4paper,fleqn,usenatbib]{mnras}

\usepackage{newtxtext,newtxmath}

\usepackage[T1]{fontenc}
\usepackage{ae,aecompl}

\usepackage{graphicx}	
\usepackage{amsmath}	
\usepackage{amssymb}	
\let\oldAA\AA
\renewcommand{\AA}{\text{\normalfont\oldAA}}
\newcommand{\teff}{T_\mathrm{eff}}
\newcommand{\logg}{\mathrm{log}\,g}
\usepackage{natbib}
\usepackage{ulem}
\usepackage{subfig}
\usepackage{float}
\usepackage[usenames,dvipsnames]{xcolor}
\usepackage{multirow}
\usepackage{footnote}
\usepackage[bottom]{footmisc}
\usepackage{ulem}
\usepackage{xltabular}
\usepackage{longtable}
\usepackage{lipsum} 
\usepackage{pdflscape}
\usepackage[export]{adjustbox}

\title[spectroscopy of M dwarfs]{First Results from MFOSC-P : Low Resolution Optical Spectroscopy of a Sample of M dwarfs within 100 parsecs}

\author[Rajpurohit, A. S. et al.]{
A. S. Rajpurohit$^{1}$,
Vipin Kumar$^{1, 2}$,
Mudit K. Srivastava$^{1}$,
F. Allard$^{3}$,
D. Homeier$^{4}$,
 \newauthor
Vaibhav Dixit$^{1}$
and Ankita Patel$^{1}$
\\
$^{1}$Astronomy \& Astrophysics Division, Physical Research Laboratory, Ahmedabad 380009, India\\
$^{2}$Indian Institute of Techonology, Gandhinagar, India\\
$^{3}$Univ Lyon, Ens de Lyon, Univ Lyon1, CNRS, Centre de Recherche Astrophysique de Lyon UMR5574, F-69007, Lyon, France \\
$^{4}$ Zentrum f\"{u}r Astronomie der Universit\"{a}t Heidelberg, Landessternwarte, K\"{o}nigstuhl 12, 69117 Heidelberg, Germany
}

\date{Accepted XXX. Received YYY; in original form ZZZ}

\pubyear{2015}

\begin{document}
\label{firstpage}
\pagerange{\pageref{firstpage}--\pageref{lastpage}}
\maketitle

\begin{abstract}

Mt. Abu Faint Object Spectrograph and Camera (MFOSC-P) is an in-house developed instrument for Physical Research Laboratory (PRL) 1.2m telescope at Mt. Abu India, commissioned in February 2019. Here we present the first science results derived from the low resolution spectroscopy program of a sample of M Dwarfs carried out during the commissioning run of MFOSC-P between February-June 2019. M dwarfs carry great significance for exoplanets searches in habitable zone and are among the promising candidates for the observatory's several ongoing observational campaigns. Determination of their accurate atmospheric properties and fundamental parameters is essential to constrain both their atmospheric and evolutionary models. In this study, we provide a low resolution (R$\sim$500) spectroscopic catalogue of 80 bright M dwarfs (J$<$10)  and classify them using their optical spectra. We have also performed the spectral synthesis and $\chi^2$ minimisation techniques to determine their fundamental parameters viz. effective temperature and surface gravity by comparing the observed spectra with the most recent BT-Settl synthetic spectra. Spectral type of M dwarfs in our sample ranges from M0 to M5. The derived effective temperature and surface gravity are ranging from 4000 K to 3000 K and  4.5 to 5.5 dex, respectively. In most of the cases, the derived spectral types are in good agreement with previously assigned photometric classification.
\end{abstract}

\begin{keywords}
Instrumentation: Spectrograph, Techniques : Spectroscopy, Stars: low-mass -- M dwarfs --Stellar atmosphere -- fundamental parameters
\end{keywords}



\section{Introduction}

Small aperture (1-2 meter class) telescopes when equipped with suitable instrumentation can broaden and diversify the scope of various science programs that could be done with such facilities.  Recent development of Mt. Abu Faint Object Spectrograph and Camera - Pathfinder (MFOSC-P) instrument \citep{Srivastava2018} for Physical Research Laboratory (PRL) 1.2m Optical-NIR telescope at Mt. Abu, India offers one such opportunity. The instrument has been developed to provide imaging and low resolution spectroscopy (R$\sim$500-2000) in optical wavelength regime along the line of widely used Faint Object Spectrograph and Camera (FOSC) series of instruments (e.g. EFOSC \citep{Buzzoni1984}, DFOSC \citep{Andersen1995} etc.) and was commissioned on the telescope in February 2019.  Though MFOSC-P has been conceived as a general user facility instrument for variety of astrophysical science programs, it can also be utilised for dedicated long term observational programs e.g. understanding the Exo-planet host stars properties. In coherence with the growing importance of Exo-planet sciences e.g. their environments, their host star properties and their dependencies,  a dedicated Exo-planet program is being carried out using purpose built PARAS spectrograph with resolution $\sim$ 67,000 \citep{Abhijit2014, Abhijit2018a} on the 1.2m PRL telescope. In coming years the program shall be further enhanced with next generation PARAS-2 spectrograph having resolution $\sim$ 100,000 \citep{Abhijit2018b} on the upcoming 2.5m PRL optical-NIR telescope at the same site. In general, M dwarf candidates are of particular interest to explore planet search in the habitable zone as they offer suitable conditions in solar neighbourhood with shorter orbital periods. Though their high resolution spectroscopy are indeed useful for detections and characterisation of host star , their intrinsic low luminosities make them difficult targets for such high resolution spectroscopy programs on small aperture telescopes. On the other hand, low resolution spectrographs like MFOSC-P can target fainter objects and provide very useful insights regarding the host star properties. Thus the feasibility of M-Dwarf characterisation program using low resolution spectra from MFOSC-P was taken as a first dedicated science program during the commissioning run of the instrument. 
\par
M dwarfs  (0.6 to 0.075 $M_{\odot}$) are the most dominant stellar component and contribute around 40$\%$ of the total stellar mass of the Galaxy \citep{Chabrier2003}. In the recent past, various space-based as well as  ground based surveys (e.g. Wide-field Infrared Survey Explorer - WISE \citep{Wright2010}, Sloan Digital Sky Survey - SDSS \citep{York2000}, Two Micron All Sky Survey - 2MASS \citep{Skrutskie2006} etc.) have been extremely useful  to provide the unprecedented photometric and spectroscopic data of cool M dwarfs. Being the suitable targets for various exoplanet search programs, a good number of bright M dwarfs candidates in the solar neighbourhood has been surveyed for their photometry \citep{lepine2011} and spectroscopy \citep{Reid1995,Hawley1996,lepine2013,frith2013}.  Survey such as Palomar/Michigan State University (Palomar-MSU) Nearby-Star Spectroscopic Survey \citep{Reid1995, Hawley1996} covers a sample of M dwarfs in both northern and southern sky beyond the 25 pc limit with radial velocities accuracy of $\pm$ 10 km/sec. This survey was used to determine spectral types, absolute magnitudes and distances of their targets, to identify chromospherically active M dwarfs with H-$\alpha$ emission and to determine the luminosity function \citep{Reid1995, Hawley1996}. Palomer-MSU survey was further used by \cite{Hawley1996} and \cite{Gizis2002} to study the relation between chromospheric activity and age among early (M0--M2.5) and mid (M3--M6) dwarfs. Later, \cite{lepine2013} and \cite{frith2013} performed the spectroscopic observations of bright M-dwarfs (magnitude J < 9 and K< 9 respectively) in the northern sky, as selected from the SUPERBLINK proper motion catalog and Position and Proper Motion Extended-L (PPMXL) catalogue. Such survey and programs provided an insight into the age, metallicity and evolution of M-dwarfs along with local star formation history.
 \par
Determination of fundamental parameters (e.g. effective temperature, metallicity, surface gravity etc.) and atmospheric properties of M dwarfs from their spectra are also of much significance. However, the atmospheric properties of M dwarfs changes significantly from early M dwarfs to late M dwarfs (M0 to M9). The presence of complex molecular bands (e.g. Titanium oxide (TiO) , Vanadium oxide (VO)  in the optical  and hydrides such as CaH, FeH in the Near-Infrared (NIR) spectra of M dwarfs etc.) make access to M dwarfs true continuum very difficult and cause uncertainties in determination of their atmospheric properties and fundamental parameters. Recently, \cite{Rajpurohit2013, Rajpurohit2014, Rajpurohit2016,Rajpurohit2018b,Rajpurohit2018a} and \cite{Passegger2016, Passegger2019} compared observed optical and NIR spectra of M dwarfs with their synthetic spectra to determine their atmospheric properties and fundamental parameters. Model atmosphere such as BT-Settl which account for recent advancement in various line list by \citep{Plez1998} and \citep{Barber2006} along with dust formation \citep{Allard2003,Allard2012,Allard2013} is now able to reproduce the shape of the spectral energy distribution (SED) down to late M dwarfs (M9) and have improved the previous estimates significantly from earlier studies \citep{Allard1995,Rajpurohit2012a}.
 \par
The principal aim of this paper is to perform the spectroscopic observations to classify and to determine the atmospheric properties and fundamental parameters of a sample of M dwarfs; thereby showing the importance of suitable instrumentation on small aperture telescopes for M Dwarfs studies and usefulness of MFOSC-P for such programs in particular. Our sample of M dwarfs along with observations and data reduction are described in Section ~\ref{sec:obs}. Section~\ref{sec:spec_class} describes the spectroscopic classification based on the comparison with the template spectra. The determination of fundamental parameters of M dwarfs in our sample is described in Section ~\ref{sec:parameters}. Section ~\ref{sec:halpha} discusses about the H-$\alpha$ emission detected in M dwarfs. In section~\ref{sec:Dis} we have discussed and summarise our results.

\section{Spectroscopic Observations and Data Reduction}
\label{sec:obs}
	
The observations were performed during the commissioning run of MFOSC-P instrument on PRL 1.2m, f/13 Telescope. We had selected bright M-Dwarf sources from the all sky catalog of bright M dwarfs \citep{lepine2011}, typically  with V magnitude brighter than 14  and covering a wide sub-spectral types. A total of 80 suitable targets were observed between February-June 2019. The authors refer to \citep{lepine2011} for more detail of the brightest M dwarfs candidates with magnitude J $\textless$ 10. The details of these targets are summurized in Table~\ref{table:table1}. Parallax for these sources are obtained from GAIA DR-2 archive \citep{Luri2018} are given as reference. These parallaxes are converted to the distance (pc) as per the method and tool given in \cite{Luri2018}. All of these sources are within the distance of 100 pc.

MFOSC-P is a fully in-house developed camera and spectrograph based on the concept of Faint Object Camera and Spectrograph (FOSC) series of instruments. The instrument provides seeing limited imaging with a sampling of 3.3 pixels per arc-seconds over the field of view of 5.2 X 5.2 square arc-minutes.  MFOSC-P uses three plane reflection gratings having 500, 300 and 150 line-pairs (lp)/mm to provide silt limited resolutions with dispersion of $\sim$ 1.1 $\AA$ per pixel, 1.9 $\AA$ per pixel and 3.8 $\AA$ per pixels respectively.  Two slits of 75 microns and 100 microns width (corresponding to 1.0 and 1.3 arc-seconds on the sky) are provided for varying seeing conditions. For more detail regarding the MFOSC-P instrument, the authors refer to \cite{Srivastava2018}. Another paper on the development, commissioning and characterization of the instrument is currently in preparation. The target M-Dwarfs in our sample were observed using 150 lp/mm grating covering the spectral range of  4800-8300 $\AA$ with a slit width of 1 arc-second.  The sources were observed for integration time in range of 500-1500 seconds per object.  Wavelength calibration spectra were recorded immediately after the each of the science spectra in the same settings of the instrument and orientation of telescope. MFOSC-P instrument is equipped with Halogen and spectral calibration lamps. Neon and Xenon calibration lamps are used for wavelength calibration. Spectro-photometric standard stars from ESO catalogue \footnote{https://www.eso.org/sci/observing/tools/standards/spectra/stanlis.html} were also observed for the instrument response correction.

\begin{figure}
\centering
  \includegraphics[height=6cm,width=0.49\textwidth]{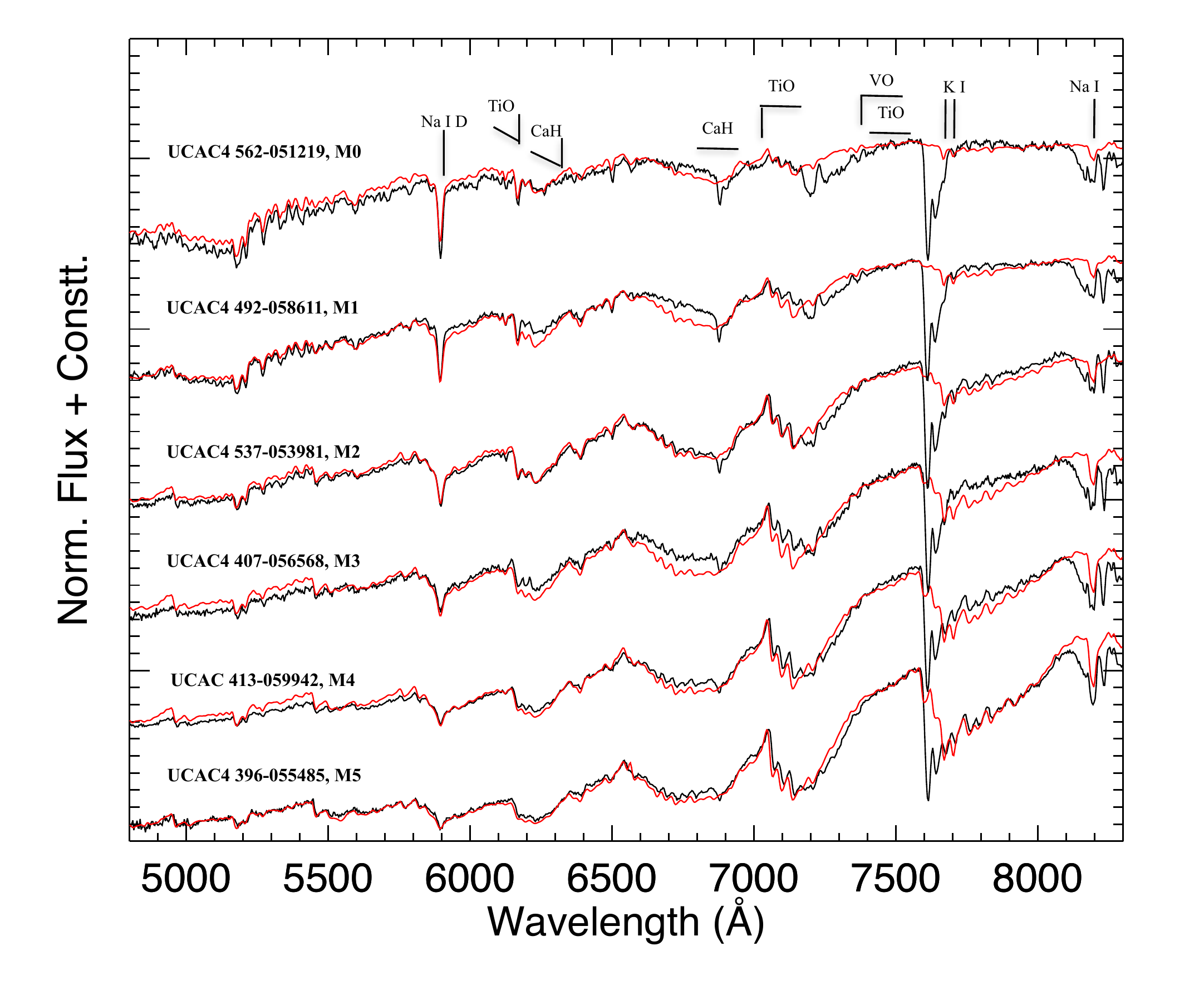} 
   \vspace{-0.2cm}
   \caption{SDSS template spectra (red) taken from \citet{Bochanski2007} is compared with observed spectral sequence of M dwarfs (black). Representative spectra of different subclasses from our sample are chosen to show the match. The most prominent spectral features along with the derived spectral type are also labeled.}
  \label{fig:fig1}
\end{figure}

\begin{figure*}
    \centering
    \includegraphics[height=9cm,width=0.40\textwidth]{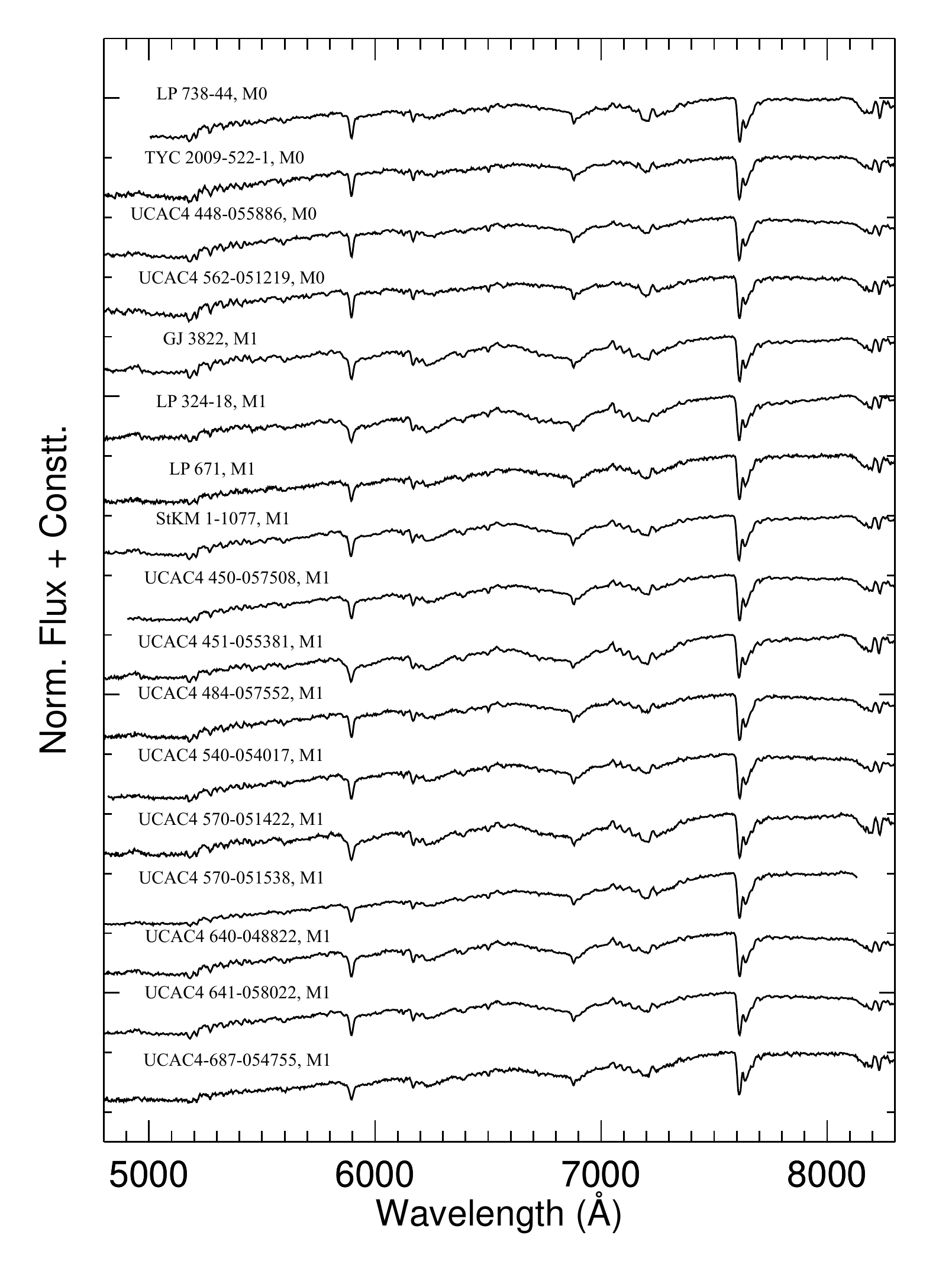}
    \includegraphics[height=9cm,width=0.40\textwidth]{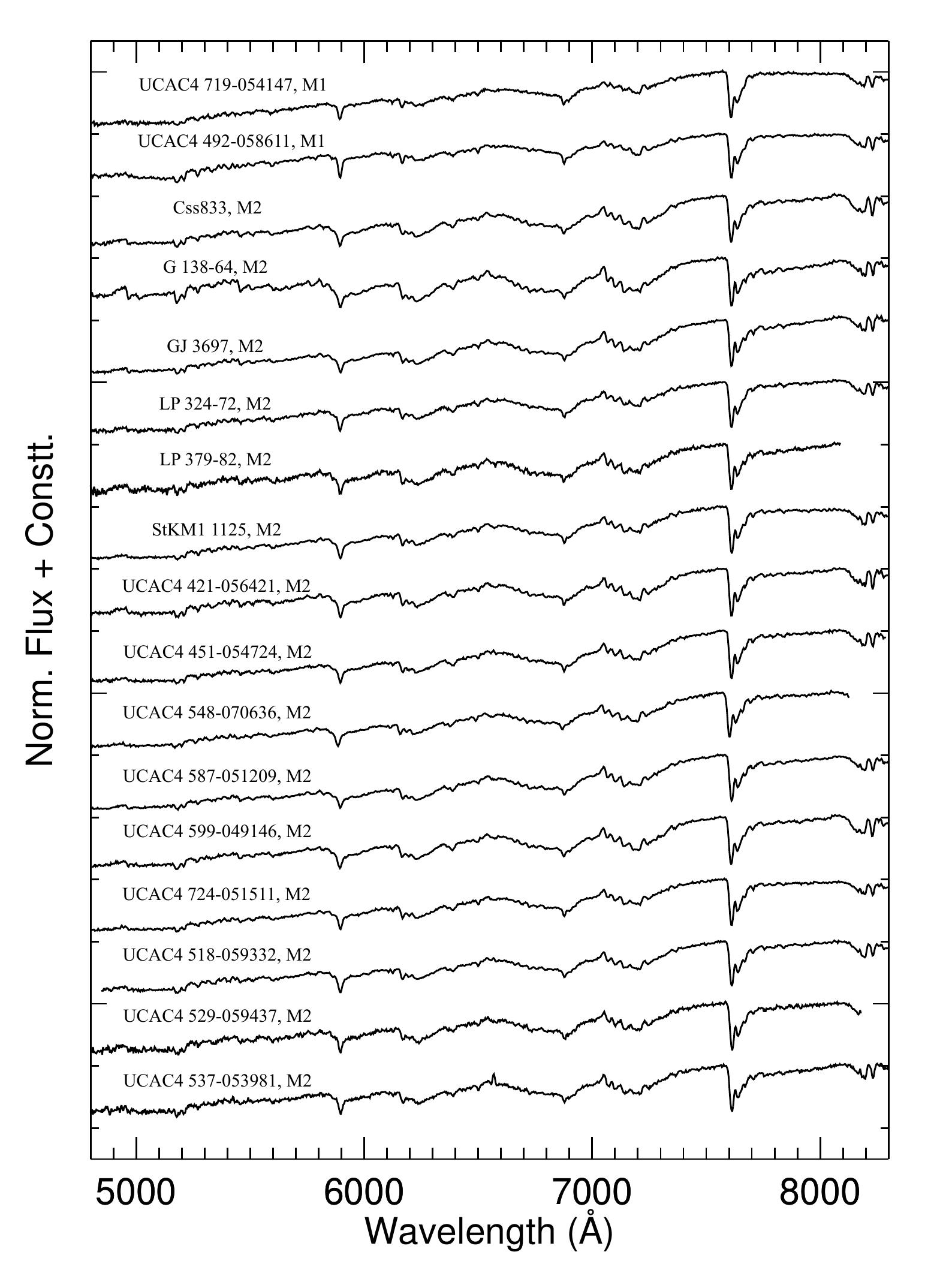}
  \vspace{-0.2cm}
    \caption{Optical spectra of M dwarfs from spectral sequence M0 to M2 observed with the MFOSC.}
  \label{fig:fig2}
\end{figure*} 

\begin{figure*}
 \centering
    \includegraphics[height=9cm,width=0.40\textwidth]{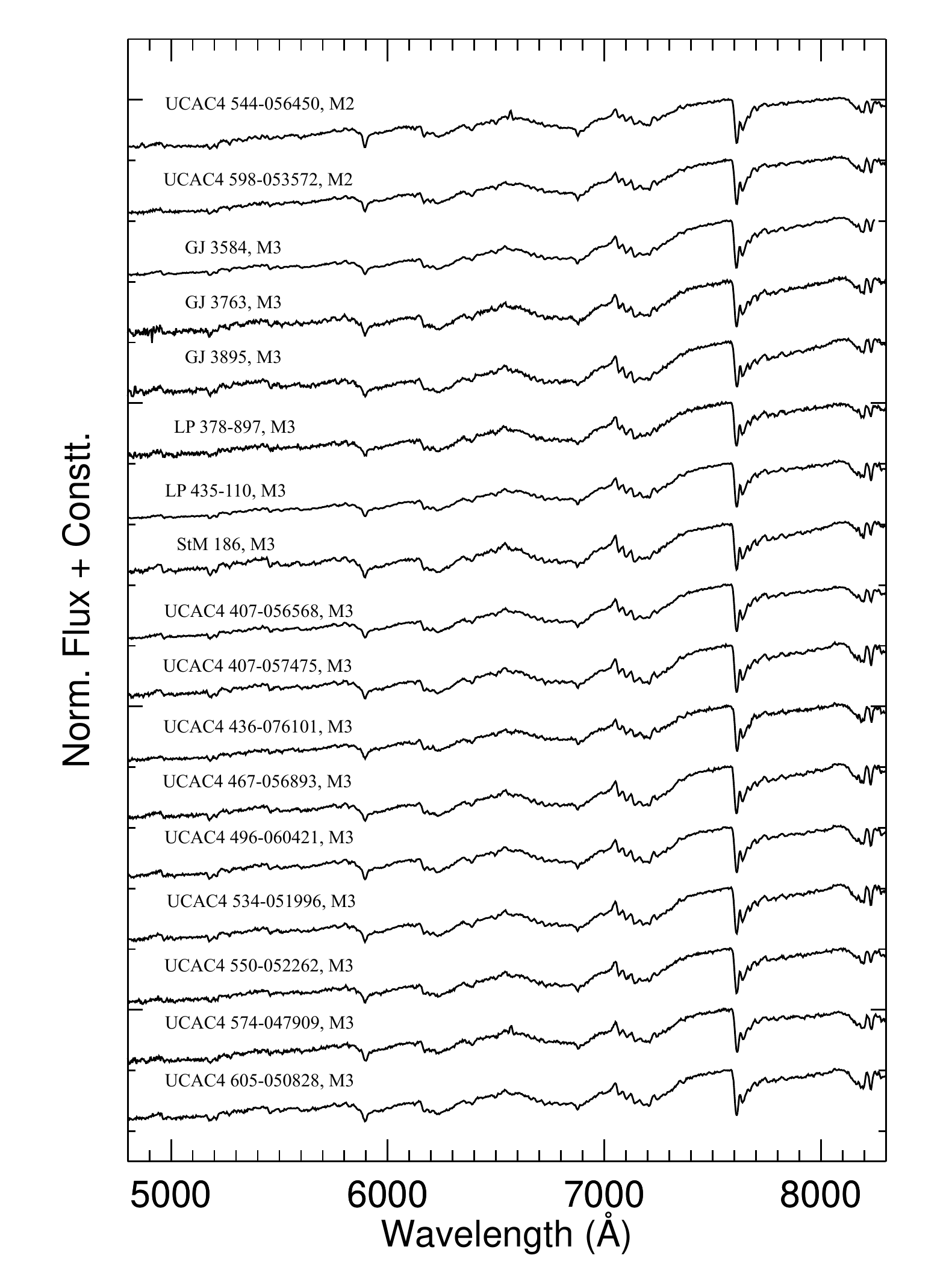}
    \includegraphics[height=9cm,width=0.40\textwidth]{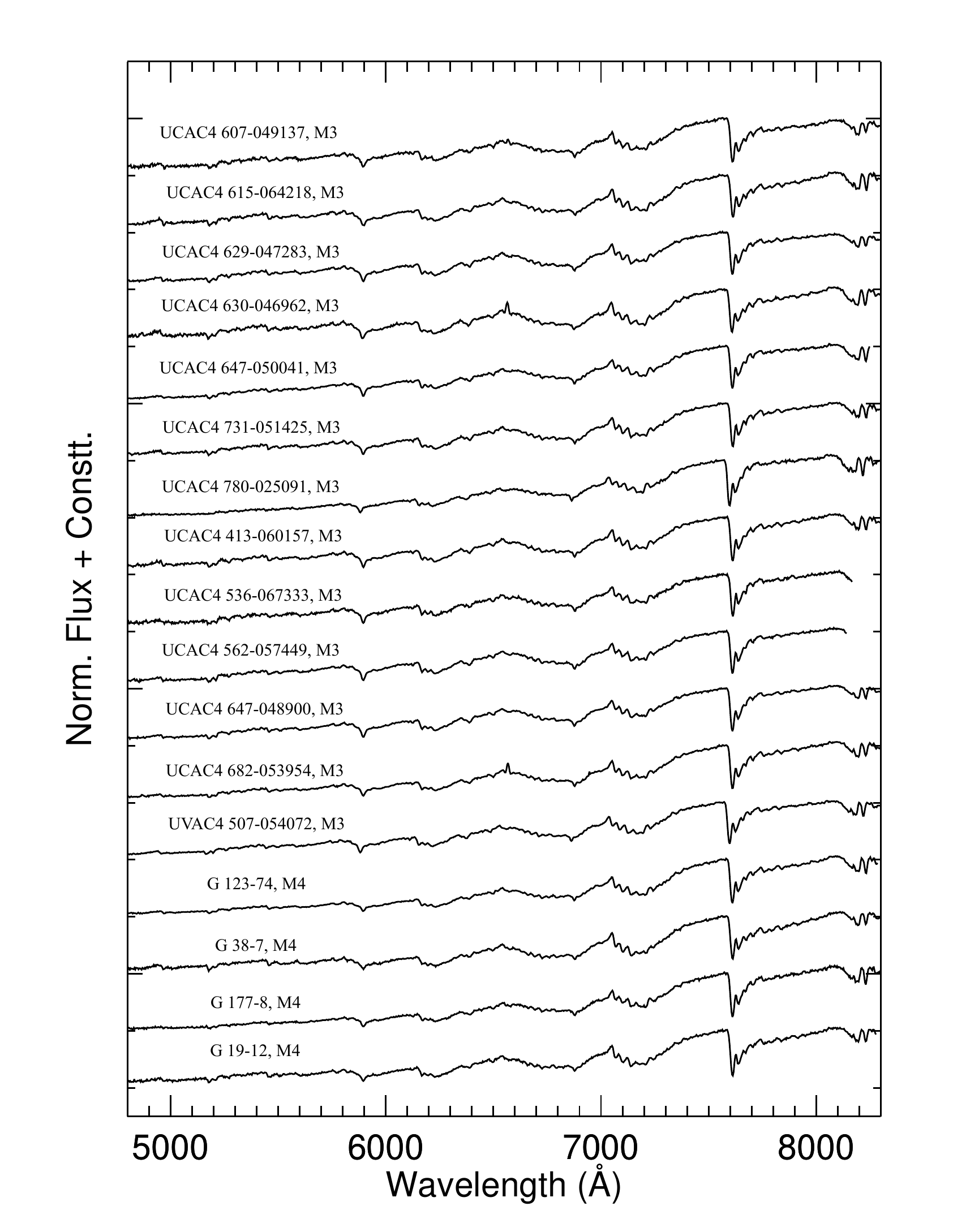}
  \vspace{-0.2cm}
   \caption{Same as fig.~\ref{fig:fig2} but for spectral sequence from M2 to M4.}
  \label{fig:fig3}
\end{figure*}

\begin{figure}
\centering
  \includegraphics[height=9cm,width=0.40\textwidth]{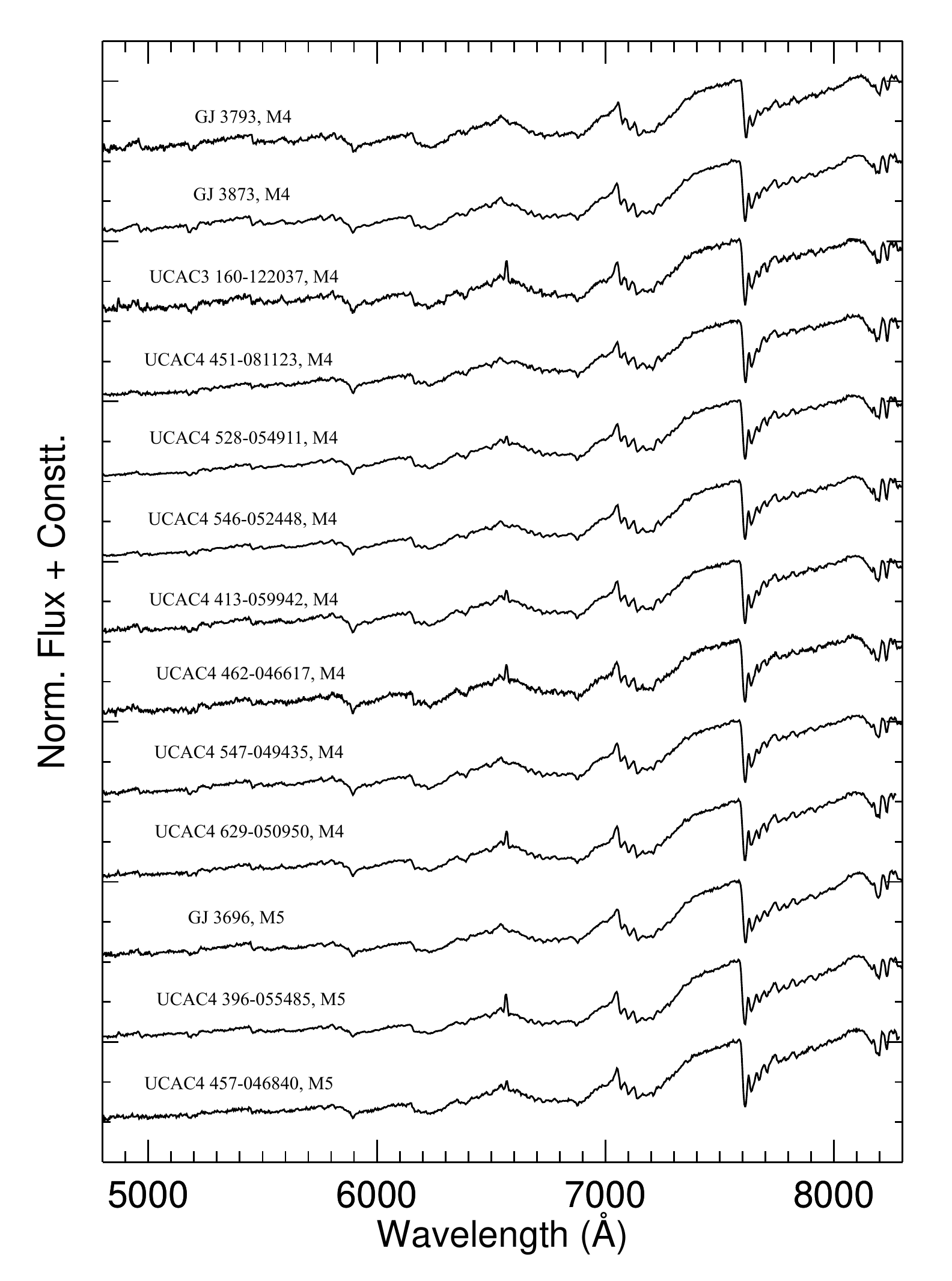}
     \caption{Same as fig.~\ref{fig:fig2} but for spectral sequence from M4 to M5.}
  \label{fig:fig4}
\end{figure}

\begin{figure*}
    \centering
    \includegraphics[height=9cm,width=0.40\textwidth]{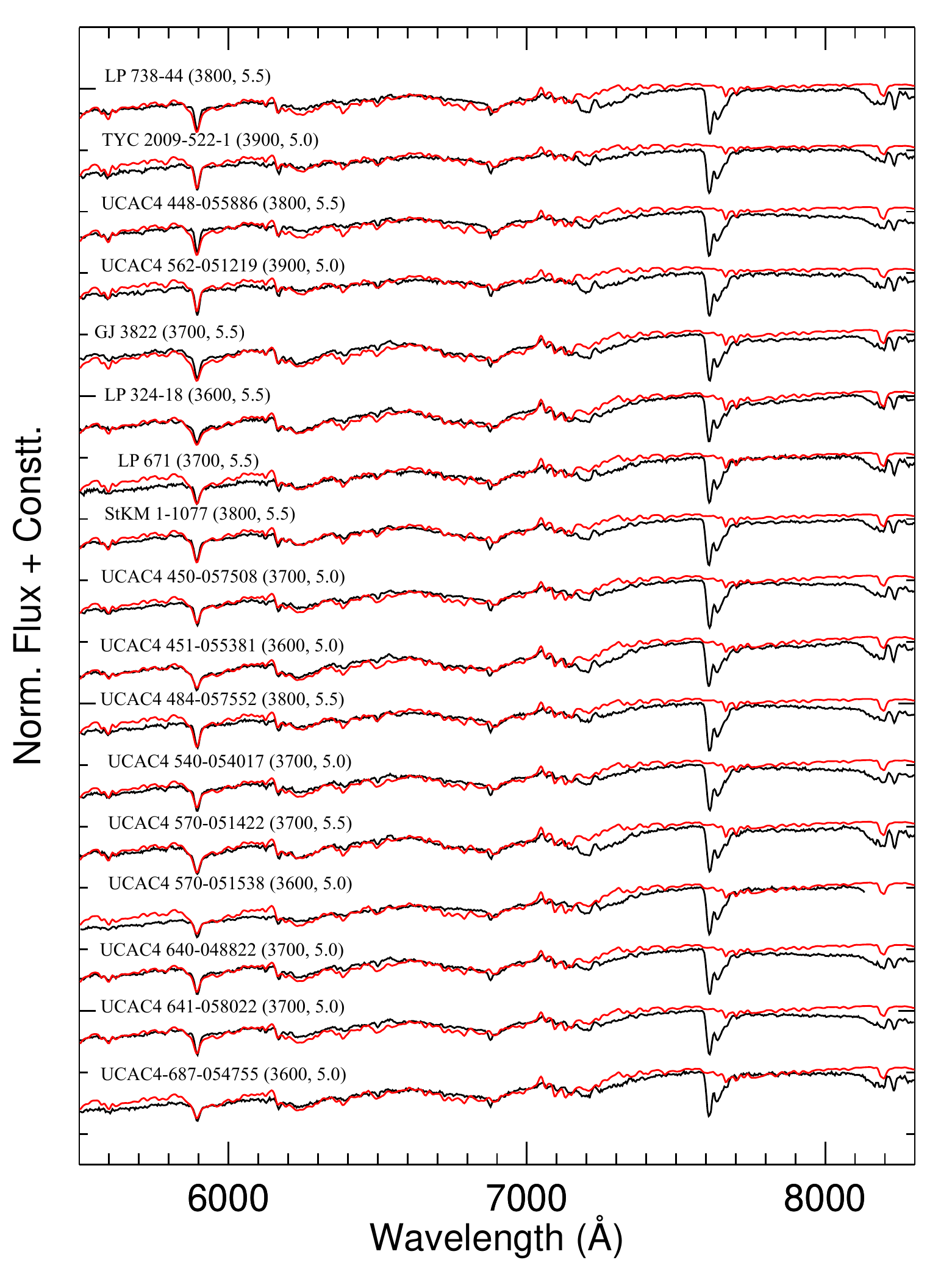}
    \includegraphics[height=9cm,width=0.40\textwidth]{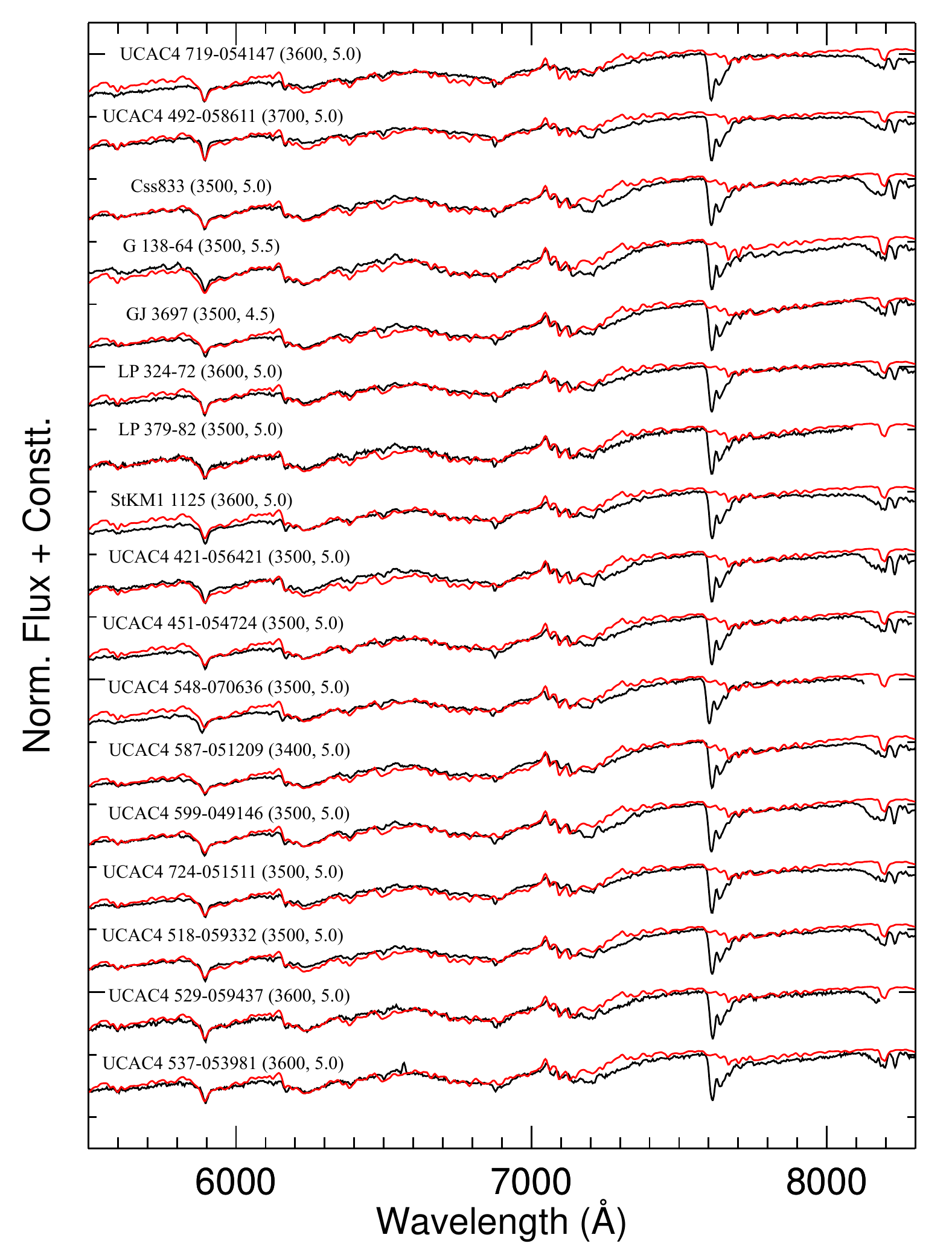}
  \vspace{-0.2cm}
    \caption{Comparison of observed spectra of M dwarfs (black) ranges from M0 to M2, with the best fit BT-Settl synthetic spectra (red). The model displayed here have $\logg$ ranges from 4.5 to 5.5 and $\teff$ ranges from 3800 K to 3400 K. Telluric features near 7600 to 7700 $\AA$ were ignored from the chi-square.}
  \label{fig:fig5}
\end{figure*} 

\begin{figure*}
 \centering
    \includegraphics[height=9cm,width=0.40\textwidth]{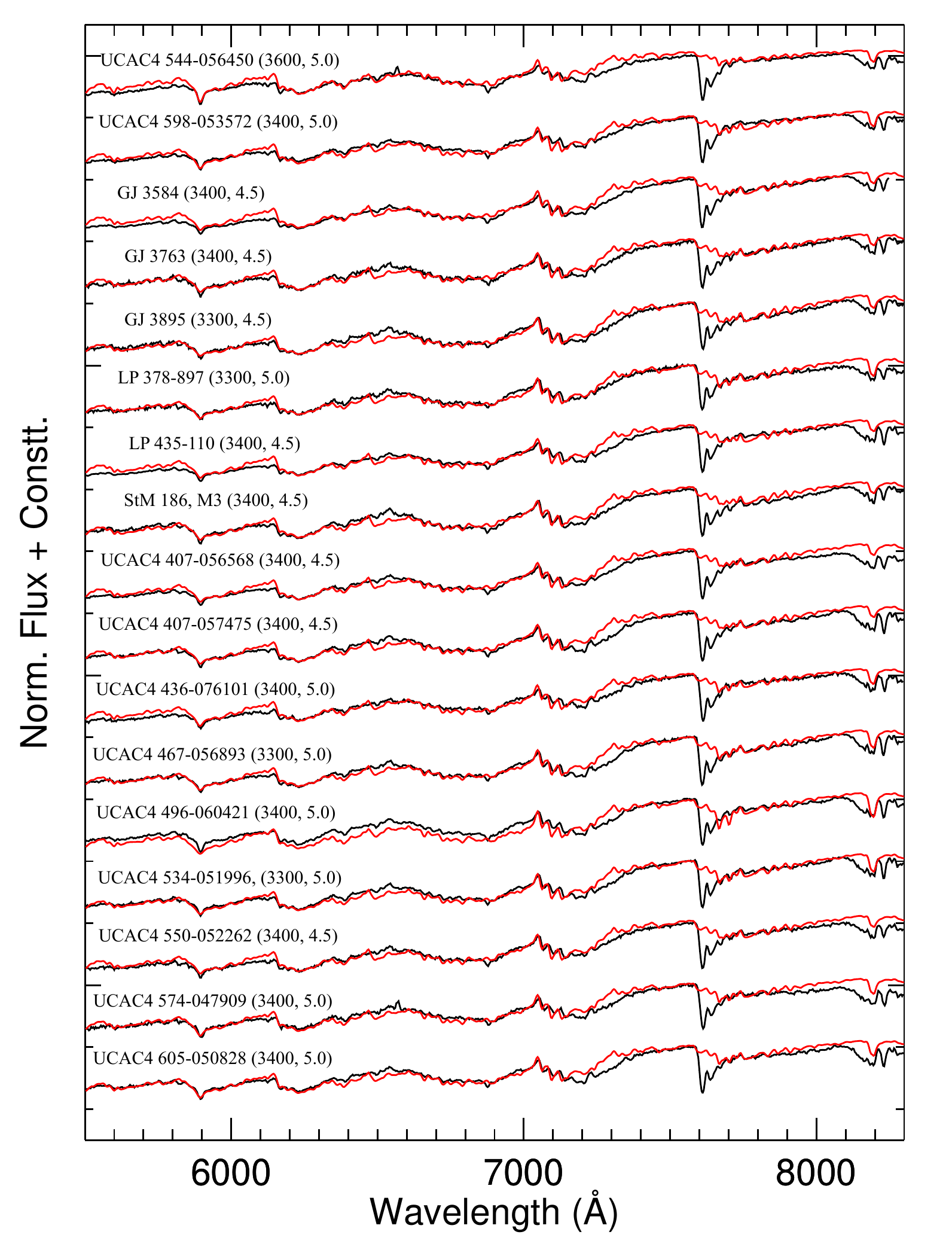}
    \includegraphics[height=9cm,width=0.40\textwidth]{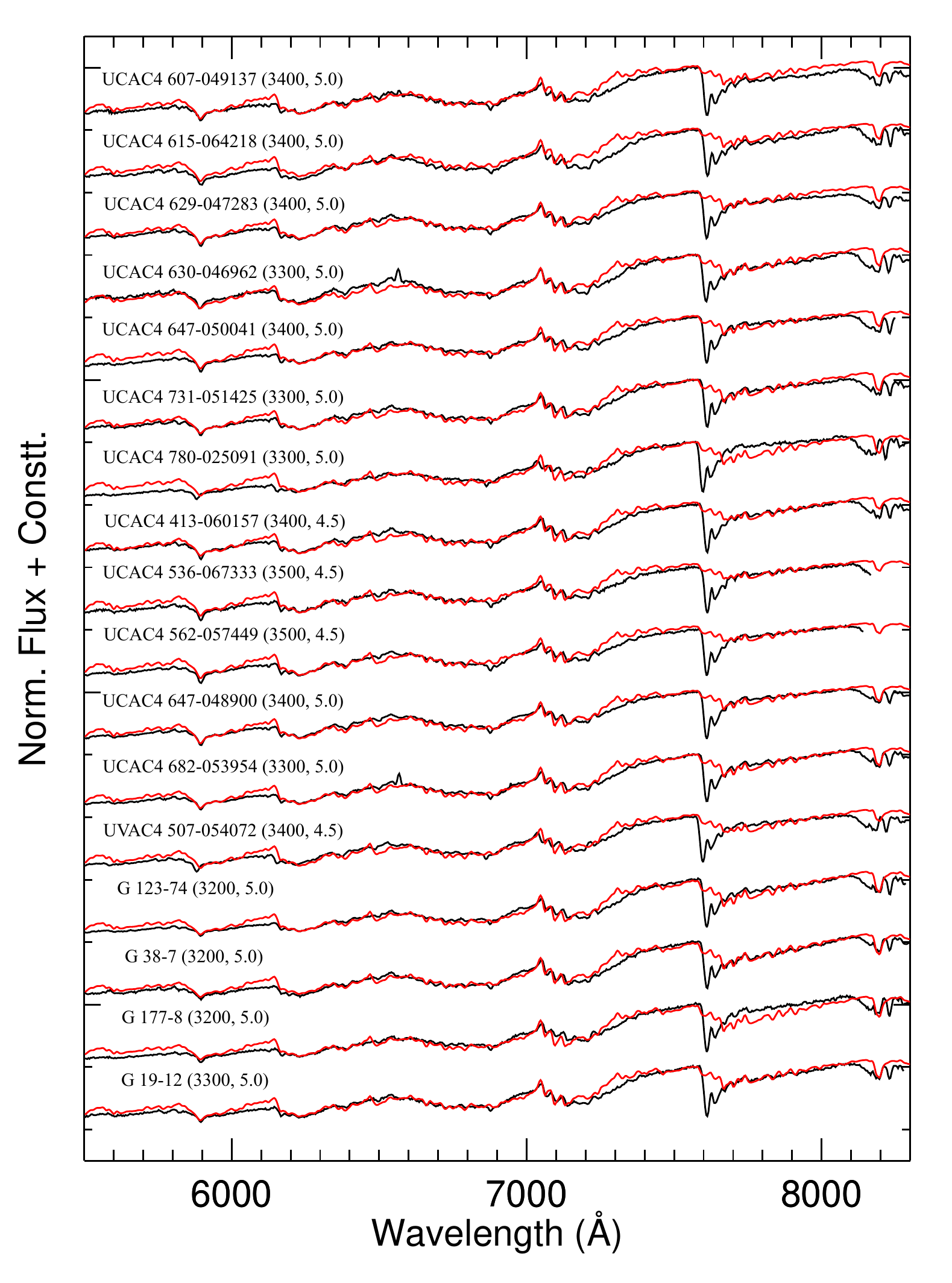}
  \vspace{-0.2cm}
    \caption{Same as fig.~\ref{fig:fig5} but for spectral type ranges from M2-M4. The model displayed here have $\logg$ ranges from 4.5 to 5.5 and $\teff$ ranges from 3600 K to 3300 K.}
  \label{fig:fig6}
\end{figure*}

\begin{figure*}
 \centering
    \includegraphics[height=9cm,width=0.40\textwidth]{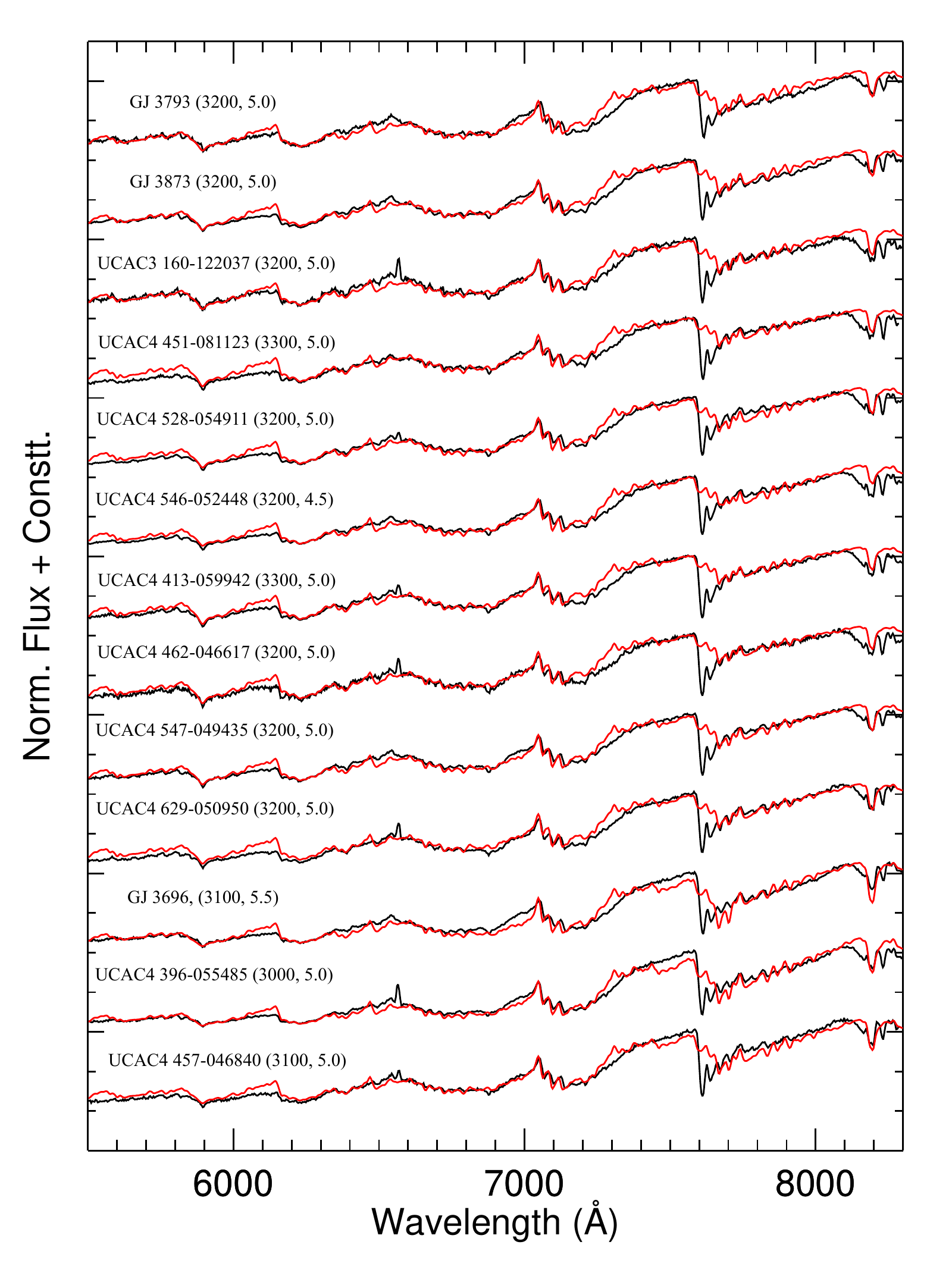}
  \vspace{-0.2cm}
    \caption{Same as fig.~\ref{fig:fig5} but for spectral type ranges from M4-M5. The model displayed here have $\logg$ ranges from 4.5 to 5.5 and $\teff$ ranges from 3300 K to 3000 K.}
  \label{fig:fig7}
\end{figure*}		

\begin{figure}
\centering
  \includegraphics[width=0.99\linewidth]{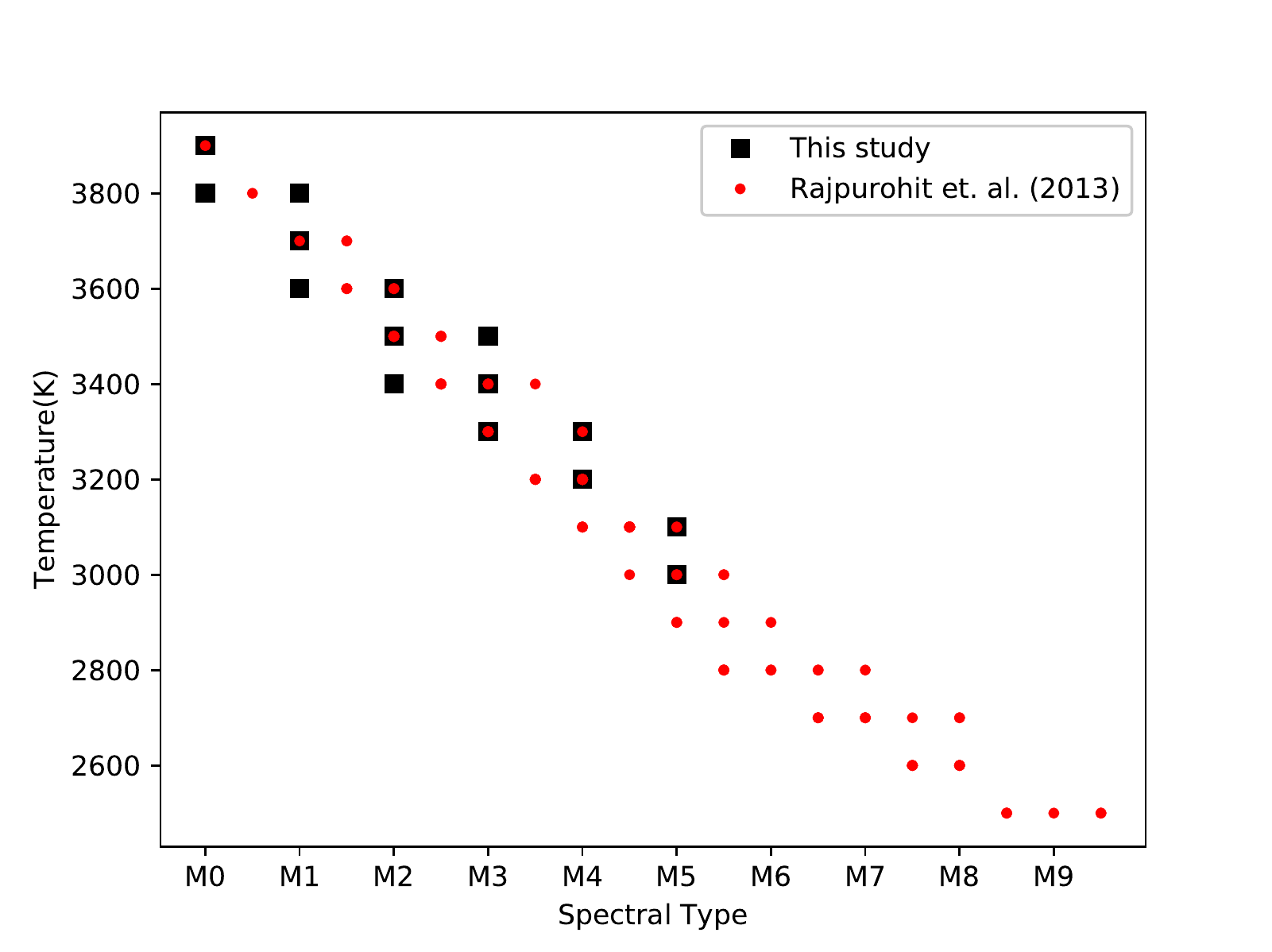}
    \vspace{-0.2cm}
  \caption{Comparison of $\teff$ versus spectral type relation from this study (black filed square) with that of \citet{Rajpurohit2013} (filed red circles) .}
  \label{fig:fig8}
\end{figure}

\begin{table*}
	\centering
	\caption{Properties of M dwarfs sample used in this study along with their coordinates. Optical and Near-infrared photometry is compiled from \citet{Ochsenbein2000} and \citet{lepine2011}. The parallaxes of M dwarfs in our sample is taken from GAIA DR-2 database \citep{Luri2018} .}
\begin{adjustbox}{width=370pt}
	\begin{tabular}{ccccccccccc}
		\hline\\
		Source & RA & Dec &V 	   & R       &    I    & J       & H       & K      &   GAIA parallex & Distance \\
		Name  &       &    & (mag)& (mag)&(mag)&(mag)& (mag)&(mag) &  (mas)  & (parsec)  \\
		\hline\\  
Css833 & 13h48m34.02s & +31d59m56.89s & 12.34 & 11.91 & \multicolumn{1}{c}{-} & 9.02 & 8.43 & 8.22 & 34.083 & 29.34 \\
G123-74 & 12h57m32.41s & +40d57m00.89s & 13.23 & 12.43 & 10.20 & 9.22 & 8.57 & 8.38 & 24.400 & 40.98 \\
G138-64 & 16h46m13.76s & +16d28m41.08s & 11.65 & 10.60 & 9.27 & 7.95 & 7.29 & 7.09 & 63.407 & 15.77 \\
G138-7 & 16h11m28.10s & +07d03m59.98s & 14.08 & 13.41 & 10.60 & 9.77 & 9.28 & 9.02 & 36.411 & 27.46 \\
G177-8 & 12h58m17.17s & +52d36m43.70s & 13.19 & 12.40 & 10.40 & 9.29 & 8.72 & 8.50 & 38.105 & 26.24 \\
G19-12 & 17h04m49.57s & +01d30m35.47s & 13.47 & 12.92 & 10.90 & 9.60 & 8.97 & 8.75 & - & - \\
GJ3584 & 10h04m32.76s & +05d33m41.25s & 12.67 & 12.22 & 9.90 & 9.02 & 8.40 & 8.18 & - & - \\
GJ3696 & 11h58m17.61s & +42d34m28.96s & 14.43 & 13.76 & 10.60 & 9.59 & 8.98 & 8.71 & 46.028 & 21.73 \\
GJ3697 & 11h58m59.45s & +42d39m39.81s & 12.09 & 11.65 & 9.90 & 8.64 & 8.05 & 7.82 & - & - \\
GJ3763 & 13h08m50.52s & +16d22m03.58s & 13.48 & 12.71 & 10.20 & 9.26 & 8.65 & 8.41 & 35.918 & 27.84 \\
GJ3793 & 13h34m49.35s & +20d11m38.67s & 14.25 & 13.42 & 10.80 & 9.67 & 9.10 & 8.85 & 18.232 & 54.85 \\
GJ3822 & 14h02m19.62s & +13d41m22.76s & 10.64 & 9.69 & 8.68 & 7.56 & 6.89 & 6.71 & 49.175 & 20.34 \\
GJ3873 & 14h54m27.92s & +35d32m56.94s & 12.55 & 12.13 & 9.50 & 8.24 & 7.71 & 7.47 & 67.072 & 14.91 \\
GJ3895 & 15h11m55.96s & +17d57m16.42s & 13.98 & 12.90 & 10.60 & 9.56 & 9.06 & 8.77 & 41.228 & 24.26 \\
LP324-18 & 13h51m45.13s & +31d42m57.67s & 13.40 & 12.46 & 10.60 & 9.59 & 8.93 & 8.75 & 14.402 & 69.44 \\
LP324-72 & 14h08m10.47s & +28d11m13.93s & 12.75 & 12.31 & 10.60 & 9.73 & 9.06 & 8.90 & 19.749 & 50.64 \\
LP378-897 & 13h16m40.56s & +23d15m42.43s & 13.51 & 13.03 & 10.70 & 9.76 & 9.11 & 8.85 & 27.695 & 36.11 \\
LP435-110 & 12h26m38.10s & +17d28m11.14s & 13.63 & 12.74 & 10.50 & 9.64 & 9.01 & 8.79 & 25.938 & 38.55 \\
LP671-33 & 10h46m07.05s & -08d22m14.77s & 12.74 & 12.32 & 11.00 & 9.86 & 9.21 & 9.03 & 17.702 & 56.49 \\
LP738-44 & 13h37m30.00s & -10d48m34.92s & 12.46 & 12.10 & \multicolumn{1}{c}{-} & 9.73 & 9.09 & 8.91 & - & - \\
StKM1-1125 & 14h08m40.58s & +23d50m54.94s & 12.34 & 11.86 & 10.20 & 9.29 & 8.62 & 8.40 & 20.267 & 49.34 \\
StKM1-1077 & 13h35m16.12s & +30d10m56.67s & 11.67 & 11.30 & \multicolumn{1}{c}{-} & 8.76 & 8.14 & 7.91 & 25.689 & 38.93 \\
StM186 & 13h41m27.65s & +48d54m45.87s & 12.98 & 12.62 & 10.20 & 9.00 & 8.45 & 8.19 & - & - \\
TYC2009-522-1 & 14h04m10.24s & +26d26m24.02s & 12.33 & 11.92 & \multicolumn{1}{c}{-} & 9.83 & 9.21 & 9.03 & 14.545 & 68.75 \\
UCAC3 160-122037 & 13h16m49.39s & -10d19m18.27s & 13.92 & 13.71 & 11.29 & 9.97 & 9.43 & 9.11 & - & 10.00 \\
UCAC4 396-055485 & 13h21m56.31s & -10d52m09.88s & 13.90 & 13.70 & 11.00 & 9.52 & 8.82 & 8.62 & 18.818 & 53.14 \\
UCAC4 407-056568 & 13h26m56.92s & -08d45m47.01s & 12.99 & 12.58 & \multicolumn{1}{c}{-} & 9.45 & 8.86 & 8.59 & - & - \\
UCAC4 407-057475 & 13h55m12.70s & -08d42m25.93s & 13.00 & 12.80 & \multicolumn{1}{c}{-} & 9.25 & 8.65 & 8.40 & 28.976 & 34.51 \\
UCAC4 421-056421 & 12h06m07.44s & -05d50m01.88s & 13.16 & 12.77 & \multicolumn{1}{c}{-} & 9.97 & 9.33 & 9.06 & 18.425 & 54.27 \\
UCAC4 436-076101 & 18h25m48.64s & -02d58m17.15s & 13.05 & 12.64 & \multicolumn{1}{c}{-} & 9.62 & 8.95 & 8.75 & 21.076 & 47.45 \\
UCAC4 448-055886 & 13h26m26.40s & -00d26m52.86s & 12.48 & 12.07 & \multicolumn{1}{c}{-} & 9.89 & 9.28 & 9.08 & 13.105 & 76.31 \\
UCAC4 450-057508 & 15h09m50.64s & -00d07m51.77s & 12.33 & 11.91 & \multicolumn{1}{c}{-} & 9.51 & 8.90 & 8.69 & 19.318 & 51.77 \\
UCAC4 451-054724 & 13h27m06.68s & +00d00m48.29s & 13.02 & 12.57 & \multicolumn{1}{c}{-} & 9.74 & 9.13 & 8.91 & 23.360 & 42.81 \\
UCAC4 451-055381 & 13h51m13.78s & +00d04m26.98s & 13.06 & 12.82 & \multicolumn{1}{c}{-} & 9.90 & 9.25 & 9.04 & 18.456 & 54.18 \\
UCAC4 451-081123 & 18h27m41.08s & +00d11m15.07s & 12.92 & 12.65 & \multicolumn{1}{c}{-} & 9.48 & 8.89 & 8.60 & - & - \\
UCAC4 467-056893 & 16h50m11.00s & +03d14m40.08s & 14.03 & 13.30 & \multicolumn{1}{c}{-} & 9.72 & 9.19 & 8.92 & - & - \\
UCAC4 484-057552 & 15h29m48.62s & +06d38m16.52s & 12.69 & 12.23 & \multicolumn{1}{c}{-} & 9.99 & 9.34 & 9.16 & 10.529 & 94.97 \\
UCAC4 496-060421 & 13h37m17.62s & +09d08m00.56s & 12.86 & 12.48 & \multicolumn{1}{c}{-} & 9.34 & 8.77 & 8.50 & 27.347 & 36.57 \\
UCAC4 528-054911 & 12h44m24.96s & +15d32m12.49s & 13.86 & 13.45 & \multicolumn{1}{c}{-} & 9.99 & 9.40 & 9.16 & 23.280 & 42.96 \\
UCAC4 534-051996 & 11h52m20.53s & +16d40m19.14s & 13.54 & 13.16 & \multicolumn{1}{c}{-} & 9.73 & 9.17 & 8.92 & 34.955 & 28.61 \\
UCAC4 540-054017 & 13h27m30.59s & +17d48m08.31s & 12.66 & 12.27 & \multicolumn{1}{c}{-} & 9.83 & 9.18 & 9.01 & 15.333 & 65.22 \\
UCAC4 546-052448 & 12h54m10.86s & +19d01m16.66s & 13.50 & 13.27 & \multicolumn{1}{c}{-} & 9.59 & 8.94 & 8.69 & 31.123 & 32.13 \\
UCAC4 548-070636 & 18h20m35.82s & +19d27m55.67s & 12.81 & 12.21 & 10.30 & 9.57 & 8.88 & 8.70 & 17.700 & 56.50 \\
UCAC4 550-052262 & 13h15m49.21s & +19d57m07.95s & 13.58 & 13.15 & \multicolumn{1}{c}{-} & 9.87 & 9.25 & 8.99 & 28.602 & 34.96 \\
UCAC4 570-051422 & 14h08m40.58s & +23d50m54.94s & 12.34 & 11.86 & 10.20 & 9.29 & 8.62 & 8.40 & 20.267 & 49.34 \\
UCAC4 570-051538 & 14h14m35.09s & +23d57m24.84s & 12.53 & 12.39 & 10.70 & 9.78 & 9.13 & 8.98 & 13.737 & 72.80 \\
UCAC4 574-047909 & 12h35m33.43s & +24d39m18.44s & 13.43 & 13.05 & 10.80 & 9.93 & 9.25 & 9.06 & 13.174 & 75.91 \\
UCAC4 587-051209 & 15h11m04.82s & +27d12m44.71s & 13.13 & 12.58 & 10.60 & 9.53 & 8.93 & 8.70 & 24.985 & 40.02 \\
UCAC4 599-049146 & 11h59m13.54s & +29d36m09.07s & 13.26 & 12.64 & \multicolumn{1}{c}{-} & 9.89 & 9.34 & 9.11 & - & - \\
UCAC4 605-050828 & 14h04m08.56s & +30d49m34.50s & 13.03 & 12.54 & \multicolumn{1}{c}{-} & 9.43 & 8.76 & 8.52 & 26.001 & 38.46 \\
UCAC4 607-049137 & 14h10m15.46s & +31d15m36.55s & 13.86 & 13.26 & \multicolumn{1}{c}{-} & 9.96 & 9.30 & 9.03 & - & - \\
UCAC4 615-064218 & 18h44m20.37s & +32d50m46.33s & 13.73 & 13.06 & \multicolumn{1}{c}{-} & 9.65 & 9.05 & 8.80 & 29.039 & 34.44 \\
UCAC4 629-047283 & 13h25m55.16s & +35d46m42.88s & 13.48 & 12.82 & 10.40 & 9.64 & 9.09 & 8.84 & 29.349 & 34.07 \\
UCAC4 630-046962 & 11h52m57.15s & +35d54m45.91s & 13.69 & 13.41 & 11.20 & 9.96 & 9.35 & 9.13 & 25.112 & 39.82 \\
UCAC4 640-048822 & 14h07m07.51s & +37d52m22.99s & 12.57 & 12.17 & 10.50 & 9.67 & 9.02 & 8.83 & 17.232 & 58.03 \\
UCAC4 641-058022 & 18h06m17.74s & +38d01m49.79s & 12.23 & 11.55 & 10.20 & 9.32 & 8.66 & 8.48 & 8.670 & 115.34 \\
UCAC4 647-050041 & 12h54m29.25s & +39d19m56.78s & 13.32 & 12.89 & 10.70 & 9.97 & 9.35 & 9.13 & 19.221 & 52.03 \\
UCAC4 687-054755 & 13h25m18.24s & +47d20m37.32s & 12.68 & 12.16 & 10.50 & 9.74 & 9.12 & 8.89 & 17.925 & 55.79 \\
UCAC4 719-054147 & 15h25m48.82s & +53d44m16.30s & 12.70 & 12.26 & \multicolumn{1}{c}{-} & 9.86 & 9.21 & 9.02 & 17.079 & 58.55 \\
UCAC4 724-051511 & 13h27m01.69s & +54d36m13.77s & 12.88 & 12.61 & 10.90 & 9.87 & 9.27 & 9.04 & 19.643 & 50.91 \\
UCAC4 731-051425 & 13h17m23.17s & +56d10m13.50s & 13.38 & 13.09 & 10.80 & 9.80 & 9.22 & 8.98 & 26.768 & 37.36 \\
UCAC4 780-025091 & 13h11m59.55s & +65d50m01.79s & 12.95 & 12.58 & 10.60 & 9.71 & 9.06 & 8.84 & 27.740 & 36.05 \\
UCAC4 413-059942 & 14h16m33.28s & -07d25m38.24s & 13.76 & 13.47 & \multicolumn{1}{c}{-} & 9.81 & 9.20 & 8.94 & 17.242 & 58.00 \\
UCAC4 413-060157 & 14h23m01.24s & -07d34m01.11s & 13.20 & 12.92 & 10.38 & 9.59 & 8.99 & 8.73 & 31.004 & 32.25 \\
UCAC4 457-046840 & 10h06m21.82s & +01d21m23.18s & 13.77 & 13.57 & \multicolumn{1}{c}{-} & 9.53 & 9.00 & 8.70 & 39.547 & 25.29 \\
UCAC4 462-046617 & 10h35m46.92s & +02d15m58.21s & 13.57 & 13.23 & \multicolumn{1}{c}{-} & 9.83 & 9.22 & 8.97 & - & - \\
UCAC4 492-058611 & 13h49m07.33s & +08d23m36.09s & 12.18 & 11.67 & \multicolumn{1}{c}{-} & 9.34 & 8.76 & 8.55 & 17.878 & 55.94 \\
UCAC4 518-059332 & 15h47m11.96s & +13d34m40.78s & 13.15 & 12.78 & \multicolumn{1}{c}{-} & 9.89 & 9.20 & 9.03 & - & - \\
UCAC4 529-059437 & 16h27m46.42s & +15d42m06.13s & 13.16 & 12.70 & 11.10 & 9.98 & 9.38 & 9.17 & 20.594 & 48.56 \\
UCAC4 536-067333 & 17h44m12.95s & +17d06m12.14s & 13.32 & 12.81 & 10.80 & 9.94 & 9.32 & 9.11 & 19.410 & 51.52 \\
UCAC4 537-053981 & 13h35m12.46s & +17d14m08.88s & 13.19 & 12.73 & 10.80 & 9.87 & 9.21 & 9.03 & 13.898 & 71.95 \\
UCAC4 544-056450 & 15h51m39.13s & +18d40m23.54s & 13.05 & 12.53 & 10.60 & 9.80 & 9.11 & 8.90 & 20.595 & 48.56 \\
UCAC4 547-049435 & 11h05m19.44s & +19d18m34.24s & 13.82 & 13.35 & \multicolumn{1}{c}{-} & 9.87 & 9.28 & 9.01 & 23.696 & 42.20 \\
UCAC4 562-051219 & 12h23m43.46s & +22d15m17.08s & 12.36 & 11.95 & 10.50 & 9.89 & 9.31 & 9.14 & 10.281 & 97.27 \\
UCAC4 562-057449 & 16h38m25.33s & +22d22m41.54s & 13.06 & 12.54 & 10.80 & 9.61 & 9.05 & 8.82 & 30.562 & 32.72 \\
UCAC4 598-053572 & 15h39m05.72s & +29d31m40.62s & 13.11 & 12.65 & \multicolumn{1}{c}{-} & 9.76 & 9.18 & 8.94 & 23.604 & 42.37 \\
UCAC4 629-050950 & 16h27m37.57s & +35d41m42.94s & 13.71 & 13.31 & 10.60 & 9.60 & 9.02 & 8.74 & 35.924 & 27.84 \\
UCAC4 647-048900 & 11h31m16.41s & +39d23m02.91s & 13.05 & 12.64 & 10.60 & 9.70 & 9.12 & 8.88 & 27.390 & 36.51 \\
UCAC4 682-053954 & 14h13m46.76s & +46d18m22.73s & 13.14 & 12.64 & 10.40 & 9.43 & 8.80 & 8.59 & 25.434 & 39.32 \\
UCAC4 507-054072 & 13h15m47.39s & +11d16m25.77s & 13.20 & 12.77 & \multicolumn{1}{c}{-} & 9.76 & 9.16 & 8.95 & 19.186 & 52.12\\
 \hline
\end{tabular}
  \label{table:table1}
\end{adjustbox}
\end{table*}

The raw data were later reduced using self developed data analysis routines in Python using astronomical image processing libraries (e.g. astropy etc.) available in public domain. The steps include bias subtraction, cosmic ray removal, tracing and extracting the spectra, sky background subtraction etc. Pixel to pixel response variations were determined using halogen spectra and were found to be less than 1$\%$, thus not applied to the observed spectra. Wavelength solution was determined from the spectral lamps spectra recorded immediately after the science observations. A third order polynomial fit was used to generate pixel versus wavelength relation. 

Though second order contamination from blue part is expected to be there at the redder part of the spectrum, the targets like M dwarfs are redder in spectrum and typically have U to I band flux ratio of nearly 1:100. Thus, given the spectral throughput of the instrument along with the telescope in blue part, blaze function of the grating and spectral energy distribution (SED) of the objects second order spectral contamination are minimal. Nevertheless, even though the observed spectral range are up to 8300 $\AA$, we have restricted our analysis within the wavelength range of 4800-8100 $\AA$ (see sections ~\ref{sec:spec_class} and ~\ref{sec:parameters}).

\section{Follow-up Spectroscopic classification}
\label{sec:spec_class}

Over the last few decades, several schemes have been proposed for M dwarfs classification. These schemes, mostly based on spectral shape and features of the M dwarf spectra, are used to preliminary classify them according to their fundamental parameters and atmospheric properties.

The spectral energy distribution (SED)  and the broadband colours of M dwarfs are mostly governed by the various molecular opacities e.g. TiO, VO and hydrides bands etc both in the optical and in the NIR. These strength of these opacities varies from early type M dwarf (M0 type) to late type (M5 of later)  for example, the  broad molecular bands such as from TiO are stronger in early M dwarfs while VO and hydrides (CaH) bands are stronger \citep{Allard2000} in later M dwarfs. The strength of these molecular bands depends on the atmospheric properties and various stellar parameters such as effective temperature ($\teff$), surface gravity ($\logg$) and metallicity ([M/H]) of the M dwarf. Considering such variation in the M dwarf spectra, \cite{Kirkpatrick1991} used the least-squares minimisation technique to classify M dwarfs by comparing the template M dwarf spectra with the target spectrum. Later on \cite{Henry2002} and \cite{Scholz2005} have used a similar technique that compares the low resolution template spectra of M dwarfs with that of the observed M dwarfs spectrum.

\cite{Reid1995} adopted the classification scheme which was based on measuring of the strength of the most prominent molecular bands called "band indices" such as TiO and CaH. Here the ratio of flux between various bandheads to that of the flux in nearby pseudo-continuum was determined which was then used to classify early M dwarfs to mid M dwarfs (M0 to M5). These bands get saturated in late M dwarfs (later than M5), thus VO bandheads were used for the classification \citep{Kirkpatrick1995}. \cite{Martin1999} assigns the spectral type to late M dwarfs based on the pseudo-continuum spectral ratios (namely PC3). \cite{Gizis1997} further classified them in to the sub category of M subdwarfs based on the strength and ratio of CaH and TiO molecular bandsheads. This work was later expanded by \cite{lepine2003b} and \cite{lepine2007}.

While the above works utilise the high resolution spectra, \cite{Scholz2005} shows that the comparison of  low resolution spectral template provides an accurate classification of M dwarfs. In this work we have utilized the low resolution spectra of M dwarfs covering the spectral regime 4800-8100 $\AA$ for their classification. The template spectra of low-mass M dwarfs are taken from \cite{Bochanski2007} to be used as template spectra of such stars from M0 to L0. These template spectra were derived from 4000 SDSS spectra. Similar to the works of \citep{Kirkpatrick1995}, \cite{Scholz2005} we have adopted the least square minimisation techniques to determine the spectral type of the observed M dwarfs in our sample.

Here we first normalised both the template and observed spectra of M dwarfs. The higher resolution (R$\sim$1800) SDSS template spectra were then convolved using gaussian kernel at the same resolution as that of the observed spectrum (R$\sim$500). Later this flux normalised spectra were compared with the template ones for the least square minimisation process to obtain the nearest match. The spectral type of this nearest match was then assigned to the observed M dwarf (Table~\ref{table:table2}). We expect the error to be of one spectral class in this method as the template spectra themselves are at the spacing of one spectral class.

Figure~\ref{fig:fig1} shows the comparison of a set of observed spectral sequence of M dwarfs in our sample with the SDSS standard M dwarfs template spectra along with the most prominent spectral features. We also tried to classify the M dwarfs in our sample by calculating spectral indices method developed by \cite{Reid1995} using the band strengths of TiO and CaH. However, the resolution of our spectra was not good enough to achieve reliable spectral types. Figure ~\ref{fig:fig2}, ~\ref{fig:fig3}, and ~\ref{fig:fig4} shows the observed spectra of M dwarfs in our sample along with their spectral type derived from the method describe above. Most of the sources shows similar spectral sub classification or within one class of their photometric classification \cite{lepine2011}. In four cases the difference is two or three subclasses.

\section{Fundamental Parameters}
\label{sec:parameters}

Fundamental stellar parameters of our sample targets were determined by comparing the observed spectra with the synthetic spectra generated by the BT-Settl version of PHOENIX \citep{Allard2010,Allard2011,Allard2013}. The BT-Settl model grid spans the $\teff$ between 300 and 7000 K in the steps of 100 K, $\logg$ ranges from 2.5 to 5.5 at a step of 0.5 dex and metallicity [M/H] ranges from -2.5 to +0.5 at a step of 0.5 dex. These models account for the latest solar abundances by \cite{caffau2009, caffau2011} with updated water vapour opacities \citep{Barber2008}. 
Various microphysical process as well as dust and cloud formation along with the gravitational settling \citep{Allard2012} has also been included in these models. Recently, \cite{Rajpurohit2012a, Rajpurohit2013, Rajpurohit2014, Rajpurohit2018a, Rajpurohit2018b} have validated the BT-Settl models by comparing the low resolution ($\triangle$$\lambda$ = 10 $\AA$) as well as the high resolution (R = 20,000 and 90,000) optical and near-infrared spectra (NIR) with the BT-Settl models in the gird range of 2400 $\le$  $\teff$ $\le$ 4000 K.

The BT-Settl model grid used in this study for the comparison spans $\teff$ between 3000 to 4000 K in steps of 100 K and $\logg$ ranges from 4.0 to 5.5 in steps of 0.5 dex. Since M dwarfs in our sample lies within 100 pc of the solar neighbourhood and belongs to the disc population \citep{lepine2011}, so we do not expect large deviations from solar metallicity. Thus we have used the models with the solar metallicity ([M/H] = 0.0) for the comparison.  The comparison of BT-Settl synthetic spectra with the observed spectra involves the process of degrading the high resolution synthetic spectra at a resolution of the observed spectra by using a Gaussian convolution. We then employed the $\chi^2$ method as discussed in \cite{Rajpurohit2013} to determine the $\teff$ and $\logg$ of M dwarfs in our sample. The spectral range between 5500 to 8100 $\AA$ have been used for the $\chi^2$ calculation. The spectral regions below 5500 $\AA$ (due to low SNR) and between 7600 to 7700 $\AA$ (which includes the telluric absorption) have   not been considered in the $\chi^2$ calculations.  During the $\chi^2$ calculations, we have not applied any weights on any spectral region for the determination of $\teff$ and $\logg$. We have retained the models that give the lowest $\chi^2$ as the best fit parameters. The best fit models have also been inspected visually by comparing them with the observed spectra. With the given resolution of the observed spectra the error in the derived fundamental parameters are equal the gird spacing of the synthetic spectrum which is 100 K for $\teff$ and 0.5 dex for $\logg$. More details about the procedure of determination of the stellar parameters of M dwarfs, can be found in  \cite{Rajpurohit2013}.  The BT-Settl model is able to reproduce the shape of SED and the profiles of the strong atomic lines such as Na I D, though no attempt has been made to fit the individual atomic lines, such as the K I and Na I resonance doublets. Figure~\ref{fig:fig5}, ~\ref{fig:fig6},~\ref{fig:fig7} shows the comparison of the entire spectral sequence of M dwarfs (black) with the synthetic spectra (red). The best fit parameters of M dwarfs in our sample is given in Table ~\ref{table:table2}. We have compared $\teff$ and spectral type determined for the individual stars in this study  with \cite{Rajpurohit2013} and found a very good agreement between them (Figure~\ref{fig:fig8} ).

\begin{table*}
	\centering
	\caption{Stellar parameters of M dwarfs sample determined in this study. }
	\resizebox{\textwidth}{!}{%
	\begin{tabular}{ccccc|ccccc}
		\hline\\
		    & Photometric & Derived & & & & Photometric & Derived & & \\
		Source & Spectral Type & Spectral Type &   $\teff$   & $\logg$  & Source & Spectral Type & Spectral Type &   $\teff$   & $\logg$   \\
		Name  & \citep{lepine2011} &  (This study)    &  (K) 	       &  (cm/sec$^2$)	& Name  & \citep{lepine2011} &  (This study)    &  (K) 	       &  (cm/sec$^2$) \\
		\hline\\ 

Css833 & M2 & M2 & 3500 & 5 & UCAC4 540-054017 & M0 & M1 & 3700 & 5 \\
G123-74 & M3 & M2 & 3500 & 5.5 & UCAC4 546-052448 & M4 & M4 & 3200 & 4.5 \\
G138-64 & M3 & M4 & 3200 & 5 & UCAC4 548-070636 & M2 & M2 & 3500 & 5 \\
G138-7 & M4 & M4 & 3200 & 5 & UCAC4 550-052262 & M3 & M3 & 3400 & 4.5 \\
G177-8 & M3 & M4 & 3200 & 5 & UCAC4 570-051422 & M1 & M1 & 3700 & 5.5 \\
G19-12 & M3 & M4 & 3300 & 5 & UCAC4 570-051538 & M1 & M1 & 3600 & 5 \\
GJ3584 & M3 & M3 & 3400 & 4.5 & UCAC4 574-047909$^\ast$ & M3 & M3 & 3400 & 5 \\
GJ3696 & M5 & M5 & 3100 & 5.5 & UCAC4 587-051209 & M2 & M2 & 3400 & 5 \\
GJ3697 & M2 & M2 & 3500 & 4.5 & UCAC4 599-049146 & M2 & M2 & 3500 & 5 \\
GJ3763 & M4 & M3 & 3400 & 4.5 & UCAC4 605-050828 & M3 & M3 & 3400 & 5 \\
GJ3793 & M4 & M4 & 3200 & 5 & UCAC4 607-049137 & M3 & M3 & 3400 & 5 \\
GJ3822 & M0 & M1 & 3700 & 5.5 & UCAC4 615-064218 & M3 & M3 & 3300 & 5 \\
GJ3873 & M4 & M4 & 3200 & 5 & UCAC4 629-047283 & M3 & M3 & 3400 & 5 \\
GJ3895 & M4 & M3 & 3300 & 4.5 & UCAC4 630-046962$^\ast$ & M3 & M3 & 3300 & 5 \\
LP324-18 & M3 & M1 & 3600 & 5.5 & UCAC4 640-048822 & M1 & M1 & 3700 & 5 \\
LP324-72 & M0 & M2 & 3600 & 5 & UCAC4 641-058022 & M0 & M1 & 3700 & 5 \\
LP378-897 & M4 & M3 & 3300 & 5 & UCAC4 647-050041 & M2 & M3 & 3400 & 5 \\
LP435-110 & M3 & M3 & 3400 & 4.5 & UCAC4 687-054755 & M1 & M1 & 3600 & 5 \\
LP671-33 & M0 & M1 & 3700 & 5.5 & UCAC4 719-054147 & M1 & M1 & 3600 & 5 \\
LP738-44 & M0 & M0 & 3800 & 5.5 & UCAC4 724-051511 & M2 & M2 & 3500 & 5 \\
StKM1-1125 & M1 & M2 & 3600 & 5 & UCAC4 731-051425 & M4 & M3 & 3300 & 5 \\
StKM1-1077 & M0 & M1 & 3800 & 5.5 & UCAC4 780-025091 & M2 & M3 & 3300 & 5 \\
StM186 & M4 & M3 & 3400 & 4.5 & UCAC4 413-059942$^\ast$ & M4 & M4 & 3300 & 5 \\
TYC2009-522-1 & M0 & M0 & 3900 & 5 & UCAC4 413-060157 & M1 & M3 & 3400 & 4.5 \\
UCAC3 160-122037$^\ast$ & M4 & M4 & 3200 & 5 & UCAC4 457-046840 & M4 & M5 & 3100 & 5 \\
UCAC4 396-055485$^\ast$ & M4 & M5 & 3000 & 5 & UCAC4 462-046617$^\ast$ & M3 & M4 & 3300 & 5 \\
UCAC4 407-056568 & M3 & M3 & 3400 & 4.5 & UCAC4 492-058611 & M0 & M1 & 3700 & 5 \\
UCAC4 407-057475 & M3 & M3 & 3400 & 4.5 & UCAC4 518-059332 & M3 & M2 & 3500 & 5 \\
UCAC4 421-056421 & M3 & M2 & 3600 & 5 & UCAC4 529-059437 & M3 & M2 & 3600 & 5 \\
UCAC4 436-076101 & M2 & M3 & 3400 & 5 & UCAC4 536-067333 & M3 & M3 & 3500 & 4.5 \\
UCAC4 448-055886 & M1 & M0 & 3800 & 5.5 & UCAC4 537-053981$^\ast$ & M3 & M2 & 3600 & 5 \\
UCAC4 450-057508 & M0 & M1 & 3700 & 5 & UCAC4 544-056450$^\ast$ & M2 & M2 & 3600 & 5 \\
UCAC4 451-054724 & M2 & M2 & 3500 & 5 & UCAC4 547-049435 & M4 & M4 & 3200 & 5 \\
UCAC4 451-055381 & M2 & M1 & 3600 & 5 & UCAC4 562-051219 & M0 & M0 & 3900 & 5 \\
UCAC4 451-081123 & M3 & M4 & 3300 & 5 & UCAC4 562-057449 & M3 & M3 & 3500 & 4.5 \\
UCAC4 467-056893 & M4 & M3 & 3300 & 5 & UCAC4 598-053572 & M3 & M2 & 3400 & 5 \\
UCAC4 484-057552 & M1 & M1 & 3800 & 5.5 & UCAC4 629-050950$^\ast$ & M4 & M4 & 3200 & 5 \\
UCAC4 496-060421 & M0 & M3 & 3400 & 5 & UCAC4 647-048900 & M2 & M3 & 3400 & 5 \\
UCAC4 528-054911 & M4 & M4 & 3200 & 5 & UCAC4 682-053954$^\ast$ & M3 & M3 & 3300 & 5 \\
UCAC4 534-051996 & M4 & M3 & 3300 & 5 & UCAC4 507-054072 & M3 & M3 & 3400 & 4.5\\
\hline
  \label{table:table2}
\end{tabular}
}
\footnotesize{$^\ast$  H-$\alpha$ emission at 6563 $\AA$ is detected.}
\end{table*}

\section{Chromospheric activity}
\label{sec:halpha}
With a thick convection zone above a radiative interior, the low mass stars in particular M dwarfs show a high level of photospheric, chromospheric, and coronal magnetic activities. In this reference, H-$\alpha$ emission at 6563 $\AA$ is of particular importance as this is an indicator of the chromospheric activity in the spectra of M dwarfs \citep{Hawley1996,Gizis2002,Reiners2008}. \cite{Hawley1996} found that late type M dwarfs i.e later than > M5 shows that incidence of activity increases monotonically as compared to early M dwarfs. They also showed that the TiO band structure depends on the chromospheric activity level of the star and thus is very useful to constrain the atmospheric model of M dwarfs. Detailed studies of such activities in field M dwarfs and M dwarfs in open clusters is of much importance as they can be used to calibrate their age activity relationship, to determine local star formation history and to understand the substellar mass function \citep{Gizis2002}. 

In our sample of 80 M dwarfs, we have detected H-$\alpha$ in 10 of them which is listed in table~\ref{table:table2}. We will be performing the follow-up spectroscopic study of the variability of the H-$\alpha$ of these objects as well as other active M dwarfs in the literature, as a proxy for the magnetic variability using MFOSC-P. Such follow-up study of these objects allow us to understand and correlate various properties such as age or rotation velocity as suggested by  \cite{Hawley1996} and \cite{Gizis2002}. 

\section{Conclusion}
\label{sec:Dis}
We have performed low resolution spectroscopic follow-up observations of 80 bright M dwarfs. All the 80 M dwarfs used in this study is having J $\textless$ 10 magnitude and are identified from the all-sky catalogue of M dwarfs \citep{lepine2011}. We have spectroscopically classified all the M dwarfs in our sample by comparing them with the standard SDDS template spectra of M dwarfs and assigned spectral type to them. The spectral type of M dwarfs in our sample are ranging from M0 to M5.  

We have used the most recent BT-Settl synthetic spectra \citep{Allard2013} to perform the spectral synthesis analysis and determined their atmospheric parameters, in particular $\teff$ and $\logg$ within an uncertainty of 100 K in $\teff$ and 0.5 dex in $\logg$. Our $\teff$ of M dwarfs ranges from 4000 K to 3000 K and $\logg$ ranges from 4.5 to 5.5 dex and is extended down to M dwarfs with spectral type M5. The recent BT-Settl model which accounts for the revise solar abundances and TiO line list is able to reproduce the SED of the M dwarfs of entire spectral sequence. 

The search for habitable planets around M dwarfs is one of the most exciting observations programs of recent times. Though in the past such programs were limited by the fewer high resolution spectroscopy facilities and faint nature of such objects, various dedicated upcoming facilities (e.g. high precision radial velocity spectrographs on larger telescopes) are expected to give a big boost to these observations programs. Such methods of high resolution spectroscopy would not only help in the detection of newer planets but also in the studies of planets and host stars environments. While the role of high resolutions spectrum is of great significance, a good deal of information about these objects can still be derived using low resolution spectra on small aperture telescopes. In this work we have attempted one such study for M dwarfs. The spectroscopic catalog presented here, along with their spectral types and stellar parameters,  is expected to provide useful list of targets for such surveys. Detection of H-$\alpha$ emission in some of the targets here presents a case to study active M dwarfs with their shorts term H-$\alpha$ variability and to understand rotation-activity relations in M dwarfs. Such programs shall be explored in future using medium resolution spectroscopy mode ($\delta\lambda \sim 2000$) of MFOSC-P.

 \section{Acknowledgments}
 MFOSC-P instrument is being funded by the Department of Space, Government of India through Physical Research Laboratory. MKS thanks to the Director, PRL for supporting MFOSC-P development program. The research leading to these results has received funding from the French "Programme National de Physique Stellaire" and the Programme National de Planetologie of CNRS (INSU). The computations were performed at the {\sl P\^ole Scientifique de Mod\'elisation Num\'erique} (PSMN) at the {\sl \'Ecole Normale Sup\'erieure} (ENS) in Lyon, and at the {\sl Gesellschaft f{\"u}r Wissenschaftliche Datenverarbeitung G{\"o}ttingen} in collaboration with the Institut f{\"u}r Astrophysik G{\"o}ttingen. DH was supported by the Collaborative Research Centre SFB 881 "The Milky Way System" (subproject A4) of the German Research Foundation (DFG). This work has made use of data from the European Space Agency (ESA) mission {\it Gaia} (\url{https://www.cosmos.esa.int/gaia}), processed by the {\it Gaia} Data Processing and Analysis Consortium (DPAC, \url{https://www.cosmos.esa.int/web/gaia/dpac/consortium}). Funding for the DPAC has been provided by national institutions, in particular the institutions participating in the {\it Gaia} Multilateral Agreement. This research has made use of the VizieR catalogue access tool, CDS,  Strasbourg, France (DOI : 10.26093/cds/vizier). The authors thank anonymous referee for his/her useful suggestions to improve the manuscript.   
\bibliographystyle{mnras}
\bibliography{ref}
\end{document}